\documentstyle[aps,epsf]{revtex}
\newcommand{\be}{\begin{eqnarray}}
\newcommand{\ee}{\end{eqnarray}}

\begin{document}
\draft

\title{\bf Collective flow in central Au--Au collisions at 150, 250 and 
400 A MeV}

\author{ {\bf Judit N\'emeth}$^1$ and {\bf G\'{a}bor 
  Papp}$^{1,2}$\thanks{Present address:
CNR, Dept. of Physics, Kent State University, Kent OH 44242, USA.}}

\address{$^1$ Institute for Theoretical Physics, E\"otv\"os University,
       H-1088 Budapest, Hungary. \\
  $^2$ GSI, Plankstr. 1, D-64220, Darmstadt, Germany.}
\date{\today} \maketitle

\begin{abstract}
Radial collective flow and thermalization are studied in gold on
gold collisions at 150, 250 and 400 A MeV bombarding energies with a
relativistically covariant formulation of a QMD code.
We find that radial flow and "thermal" energies calculated for all the
charged fragments agree reasonably with the experimental values.
The experimental hardware filter at small angles used in the FOPI
experiments at higher energies selects mainly the
thermalized particles. 
\end{abstract}
\pacs{{\em PACS numbers:\/} 25.75.Ld, 25.75.-q, 25.70Mn\\
  \noindent {\em Keywords:\/} Collective flow; Molecular Dynamics}


In recent years central-collisions studies became a focus of attention in
the intermediate energy domain (100-500 A MeV)~\cite{FOPI,EOS}.
One of the measurables concentrated on by the experiments is connected to
the flow.
It is a well-known fact~\cite{PREDICT},
that at larger impact parameters there is a
sideward flow and a squeeze-out flow; these quantities were
measured~\cite{FLOWEXP} 
and calculated~\cite{DAN}. The sideward and squeeze-out
flows mostly disappear in very central collisions. However,
calculations~\cite{DAN} predicted a large collective energy (radial
flow) in central
collisions and this was confirmed later by experiments~\cite{FLOW}.
This collective 
energy can be visualized with a blast model~\cite{FOPI,BLAST},
where the system expands spherically around the center of mass.
\begin{figure}[htbp]
\centerline{\epsfysize=45mm \epsfbox{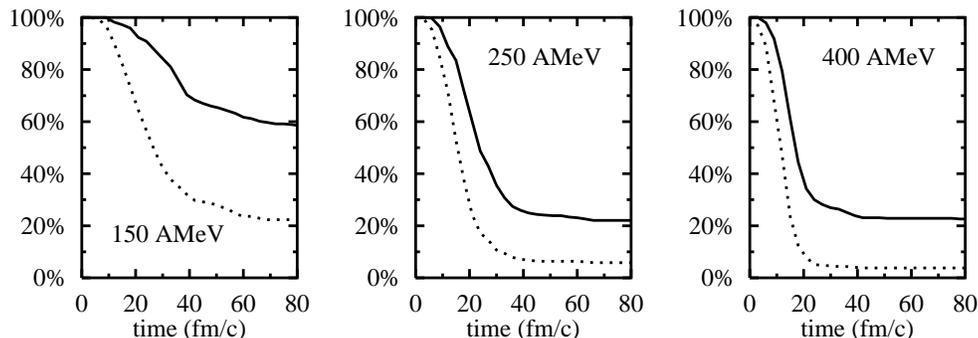}}
\caption{Number of particles that collided less than three times (solid
line), and without collisions (dotted line) for three incident
energies, at $b=0.5 fm$ impact parameter, as the function of time. Initially 
($t=0$) the nuclei are at a distance of 2 fm from each other and the
total overlap for free evolution would occur at 29, 23 and 19 fm/c for
the three energies, respectively.}
\label{fig-cold}
\end{figure}

Recently we deduced a momentum dependent, relativistically
invariant two-body force~\cite{USHIP}, which can be applied to QMD 
calculations~\cite{AICH}. To check the validity of the force,
we made detailed
calculations for central gold on gold collisions in the energy domain
150--400 A MeV~\cite{USHIP} and compared the results with experimental
data~\cite{FOPI}. The agreement turned out to be highly satisfactory;
even the number of the intermediate mass fragments (IMF) is very close
to the experimentally measured value. It is worth examining what
can we learn about the radial flow using this model.

First we study to what extent one may speak about thermalization of
the nucleons in central and semi-central collisions. The system is
usually assumed to be thermalized if the nucleons collided a few ($\sim
3$) times.
From Fig.~\ref{fig-cold} one can see that for higher energies ($250-400
A MeV$) only 20\% of the nucleons did not collide more than twice (and
may be considered as unthermalized). At low energy (150 A MeV) this
fraction amounts to 60\%, due to the large Pauli blocking (for 150
AMeV 65\% of the possible collisions were blocked, while for 400 AMeV
the blocking was only 25\%).  We followed the path of some
nucleons in collisions with 400 A MeV energy. Those nucleons which did
not collide at all (less than 3\%) are generally positioned in the outer
layer of the colliding nuclei.

\begin{figure}[htbp]
\centerline{\epsfysize=45mm \epsfbox{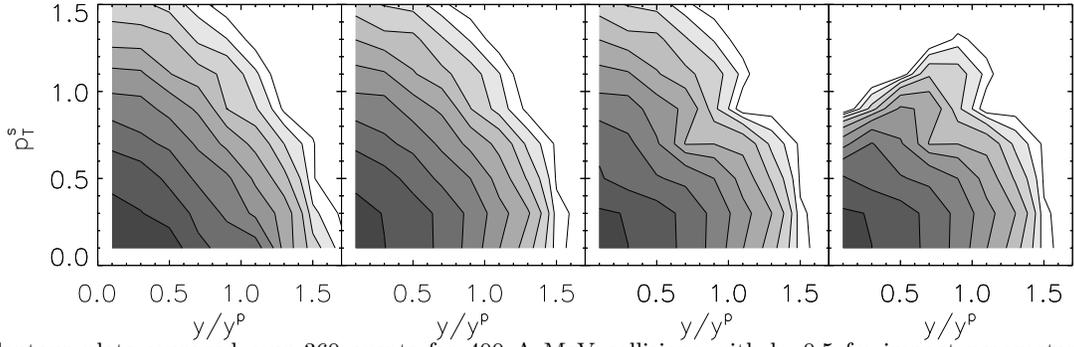}}
\caption{Contour plots averaged over 360 events for 400 A MeV collisions
with 
b=0.5 fm impact parameter. Shown are the
raw proton distribution (far left), the distribution with a
$\theta_{lab}>1.2^\circ$
experimental hardware filter
(center left), with an additional $\theta_{lab}\sim 21^\circ$ filter (center
right), and with an extra $\theta_{lab}<30^\circ$ filter (far right) in the
scaled momentum -- rapidity plane.
The hardware filter around $\theta_{lab}\sim5^\circ$ is not visible. The
contours are separated by factors 1.5.}
\label{fig-contc4}
\end{figure}
Assuming that nucleons having collided at least
three times are thermalized, one may conclude
from Fig.~\ref{fig-cold}, that at least for high energies
the thermalization rate is $\sim 80\%$. For a spatially homogeneous
system this would mean an isotropic momentum distribution.
For this reason it is interesting to examine the momentum distribution
by making a contour plot of the invariant cross section
$d^2\sigma/p_Tdp_Tdy$ of the outgoing single protons in the two dimensional
space of the transverse momentum scaled with the projectile momentum,
$p_T^p$, and the rapidity $y$ scaled with the projectile
rapidity $y^p$ 
in the center of mass system. We note that in the energy domain of
interest (up to $600MeV$) the scaled rapidity and scaled momentum are
equal to each other within 5\%. As a consequence, a thermalized
distribution should 
appear as circles in the $y-p$ plane. First we examined the contour
plots without
experimental filters at $b=0.5 fm$ and 400 A MeV collision energy after 150
fm/c evolution. One can
see from Fig.~\ref{fig-contc4}, that in contrast to a thermalized
system, the raw data do not show an isotropic distribution (see the
most inner -- highest multiplicity contour, having a distortion factor
$\sim 1.5$).
However, applying {\em only} the seemingly negligible small laboratory
forward angle filter (excluding particles at $\theta_{lab}<1.2^\circ$),
which means the 
exclusion of a small number of protons only, a nearly isotropic
distribution is recovered. The interpretation of this result is the
following: the unthermalized (not yet collided) protons leave the
collision region with high
energy in the forward direction; thus, the remaining proton distribution
is already thermalized.
Applying further angular filters does not change this behavior.
\begin{figure}[htbp]
\centerline{\epsfysize=45mm \epsfbox{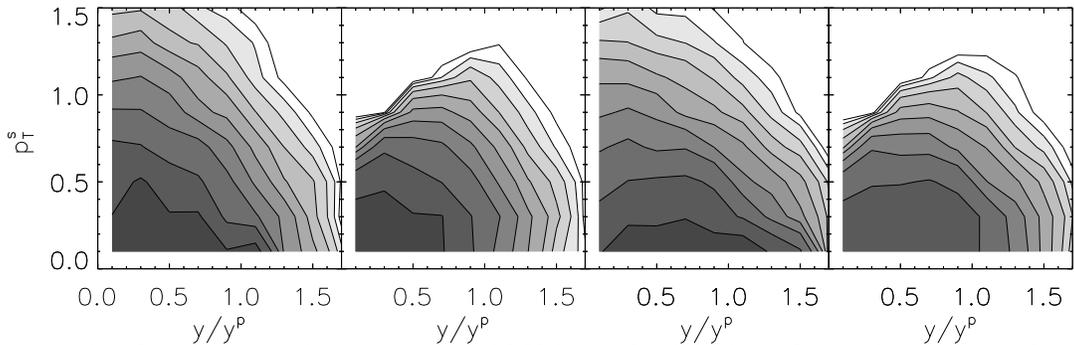}}
\caption{Contour plots for proton distributions at 150 A MeV, b=1.5 fm 
(left block of 2 panels) and b=7.5 fm (right block of 2 panels) in the
scaled momentum -- 
rapidity plain. The left figures of each block show the raw data, while the 
righthand ones show the data
with the most relevant $\theta_{lab}>1.2^\circ$ and
 $\theta_{lab}<30^\circ$ hardware filters.}
\label{fig-cont1}
\end{figure}

\begin{figure}[htbp]
\centerline{\epsfysize=45mm \epsfbox{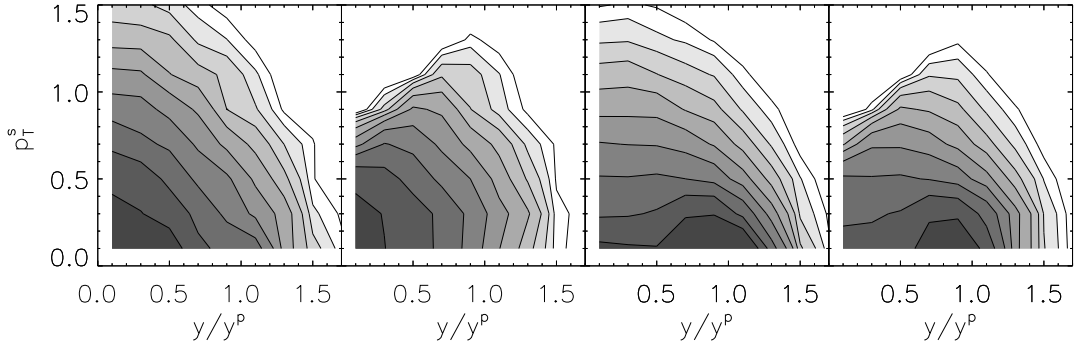}}
\caption{Same as Fig.~\protect\ref{fig-cont1} at 400 AMeV.}
\label{fig-cont4}
\end{figure}

Figs.~\ref{fig-cont1} and~\ref{fig-cont4} show contour plots
for protons at 150 and 400 A MeV, respectively, for central and
peripheral events. In order to compare our result with the experimental
one, where 
the effect of the filters in the range $1.2^\circ<\theta_{lab}<
30^\circ$ was averaged out, we took only the relevant
$1.2^\circ<\theta_{lab}$ and $\theta_{lab}< 30^\circ$ filters. One can
see that for central events the filtered distributions are almost
isotropic (except for the $\theta_{lab}< 30^\circ$ filters), however, the raw,
unfiltered values are not.
This fact is very pronounced in the case of 150 A MeV.
In the case of large impact parameters the
distributions are always distorted. The same observation was made in
connection with the experimental measurements~\cite{FOPI}.

Experimentally the collective radial flow is determined from the
kinetic energy distribution of the large fragments. 
However, since at 400 MeV the total mass of 
all large fragments for central collisions is less than 3\% of all
the particles, such a method produces very poor statistics in our
case. In order to determine the radial flow we considered to
possibilities:  calculating the flow in the 1. {\bf gas scenario}, when
all the nucleons are considered to be individual objects contributing to
the radial flow; or in the 2. {\bf cluster scenario}, when all the
charged fragments (single protons and clusters) are taken into
account. For comparison we present results for both.

In order to determine the radial flow of the nucleons we divided the
solid angle into 32 equal pieces and calculated the flow velocity 
$\beta_{(k)}$ within each sector $k$ ($k$=1,32) as
\be
  u_{(k)} = \sum_{i\in\Omega_k}
	\frac{\vec{r}_i\vec{p}_i}{|\vec{r}_i|} \mbox{\Large /}
	\sum_{i\in\Omega_k} m_i
		\quad , \quad\,
  \beta_{(k)} = \frac{u_{(k)}}{\sqrt{1+u_{(k)}^2}} \,
\ee
where $u_{(k)}$ is the four velocity associated to sector $k$ and the
summation is extended for all the charged particles within the solid
angle $\Omega_k$, with $m_i$ being the mass of a fragment, while $r_i$
is the position of the charged particle at the end of the calculation
(150 fm/c). We repeated the same for the gas scenario with each $m_i$ being
the nucleon mass.

The fluctuation of the flow velocity is less than 2\% in all the 32
sections for each energy at $b=0.5 fm$ both for the
cluster and gas algorithm. Furthermore, the fluctuation of the number of
particles within the sectors is at most 3-4\%. The extracted values are
also stable against changing the end-time of the calculation: from the
freeze-out time up to twice the freeze-out time the change of the flow
energy is 10\% and does not change further for larger end-times. 
These results show that the radial flow can be determined
in a very reliable way.
Encouraged by the isotropy of the system and by the
lack of other collective excitations (side flow or squeeze-out) we
define a disoriented ("thermal") energy as the rest kinetic 
energy\footnote{The energy
defined here is {\em not} exactly the thermal energy, since a thermal
energy can be defined for completely thermalized systems only in the local
rest frame. As the complicated experimental procedure obtaining the
thermal energy cannot be reproduced within the given model, we consider
the disoriented energy $E_{do}$ to be close to the thermal one.},
\be
  E_{do}&=& \frac 1{\displaystyle \sum_{i=1}^M A_i}
	 \sum_{i=1}^M \left( 
	\sqrt{m_i^2+p_i^2}-m_i \right) -E_{flow}
	\quad \mbox{with} \nonumber \\
  E_{flow}&=& \frac 1{\displaystyle \sum_{i=1}^M A_i}
	 \sum_{i=1}^M \left(
	\sqrt{m_i^2+ 
	\frac{\left(\vec{r}_i\vec{p}_i\right)^2}{r_i^2}}-m_i\right) \,
\ee
where $A_i$ is the mass number of a cluster, $M$ is the total
multiplicity of the clusters (and $M=A$, the
total number of particles, for the gas algorithm).

In Table~\ref{tab-vel} we give the
calculated flow velocity, the flow energies, and the "disoriented/thermal" 
energies
evaluated both for the gas and for the fragments (cluster). As a
comparison, we give the values extracted
from Ref.~\cite{FOPI}, where they use a blast wave model fit to the
experimentally measured kinetic energies of heavy fragments. We note
that these quantities are model dependent, and evaluating them we used
the natural way for the QMD model, not the experimental procedure. We find
the agreement surprising.

Finally we make a remark comparing our result to the one of the
EOS~\cite{EOS} group. In this experiment the selection criteria were
based on the multiplicity only and the deduced values of the radial flow
velocity are considerably lower compared to the flow velocities of the
FOPI experiment. In our model we find that the multiplicity trigger is
not sufficient enough for selecting the central collisions; we got
a considerable amount of events contributing from $b=4.5-5 fm$. As a
result, using the multiplicity trigger, our flow velocities, $\beta$, are
reduced by $\sim 15-20\%$ in agreement with the EOS result. Consequently
the ``thermal'' energy is increasing by the same amount.

In conclusion, we have investigated collective radial flow in the case of
Au+Au collision at 150, 250 and 400 A MeV. The results were compared
to FOPI experiments~\cite{FOPI} and we found a reasonable
agreement. Furthermore the experimental setup automatically
filters out the unthermalized particles for higher energies. These
results suggest that the compressibility and the 
momentum dependence of the used force is highly satisfactory. The
application of our force for higher energies when particle creation is
important, is in progress.
\begin{table}
\caption{Average flow energies (MeV), ``disoriented/thermal'' energy and flow
velocity at 400 (upper), 250 (middle) and 150 A MeV (lower part) at b=1
fm impact parameter for the gas (g) and cluster (c) assumption. 
}
\label{tab-vel}
\begin{center}
\begin{tabular}{||r||c|c||c|c||c|c||}
 & E$_{\rm flow}$ & E$^{exp}_{\rm flow}$
	& E$_{do}$ & E$^{exp}_{th}$ & 
		$\beta$ & $\beta^{exp}$ \\[1mm] \hline
g (400)& 63.2$\pm$2.5& - & 37.2$\pm$3 & - & 0.345$\pm$0.006 & - \\
c (400)& 59.7$\pm$2.5 & 56.8$\pm$6.3 & 32.7$\pm$2 & 32.8$\pm$6.3& 
	0.338$\pm$0.006 & 0.334$\pm$0.017 \\ \hline
g (250)& 41.2$\pm$1.2 & - & 18.2$\pm$0.5 & - & 0.285$\pm$0.002 & - \\
c (250)& 34.8$\pm$1.0 & 34.0$\pm$3.9 & 16.8$\pm$0.6 & 21.5$\pm$3.9 & 
	0.264$\pm$0.002 & 0.263$\pm$0.014 \\ \hline
g (150)& 21.6$\pm$0.5 & - & 12.9$\pm$0.8 & - & 0.21$\pm$0.002 & - \\
c (150)& 18.3$\pm$0.3 & 19.9$\pm$2.3 & 10.2$\pm$0.5 & 12.6$\pm$2.3 & 
	0.194$\pm$0.002 & 0.204$\pm$0.011
\end{tabular}
\end{center}
\end{table}

\vglue 0.2cm
{\bf \noindent Acknowledgments \hfil}
\vglue 0.4cm

We would like to thank to Prof. H. Feldmeier for the continuous
discussions and suggestions and to Prof. G. Fai for carefully reading
the manuscript.
One of the authors (J.N.) should like to express her thanks to
Prof. W. Greiner and the University of Frankfurt and to
Prof. W. N\"orenberg and the GSI for their kind hospitality, during her
visits to these institutions, where part of 
this work was done. Discussions with W. Reisdorf and A. Gobbi
are highly acknowledged. We express our thanks to the FOPI group
providing the experimental filter program.
This work was partly supported by 
Hungarian OTKA grant T022931 and 
FKFP grant 0126/1997.

\end{document}